\documentclass[letterpaper,journal]{IEEEtran}
\PassOptionsToPackage{table}{xcolor}
\usepackage{amsmath,amsfonts}
\usepackage{algorithmic}
\usepackage{algorithm}
\usepackage{array}
\usepackage[caption=false,font=normalsize,labelfont=sf,textfont=sf]{subfig}
\usepackage{textcomp}
\usepackage{stfloats}
\usepackage{url}
\usepackage{verbatim}
\usepackage{graphicx}
\usepackage{cite}
\usepackage{svg}
\usepackage{multirow}
\usepackage{makecell}
\usepackage{bbding}
\usepackage{tabularx}
\usepackage{booktabs}
\usepackage{listings}
\usepackage{amsthm}
\usepackage{xcolor}
\usepackage{hyperref}
\usepackage{pdfpages}

\definecolor{caseid}{RGB}{215,232,248}
\definecolor{caseloop}{RGB}{218,237,211}
\definecolor{casecond}{RGB}{255,226,208}
\definecolor{caseupdate}{RGB}{232,222,246}

\begin{document}

\title{RoCoMark: Robust Code Watermarking Against LLM-Based Functionality-Preserving Rewriting Attacks}

\author{Hongjie Zhang, Xingcheng Li, Hanzhou Wu, Zhihua Xia and Xinpeng Zhang\thanks{H. Zhang and X. Li are with College of Computer Science, Sichuan Normal University, Chengdu 610066, China (E-mails: zhanghongjie@sicnu.edu.cn, secndzz7@gmail.com).}\thanks{H. Wu is with the School of Communication and Information Engineering, Shanghai University, Shanghai 200444, China (E-mail: h.wu.phd@ieee.org).}\thanks{Z. Xia is with the College of Cyber Security, Jinan University, Guangzhou 511443, China (e-mail: xia\_zhihua@163.com).}\thanks{X. Zhang is with the School of Computer Engineering and Science, Shanghai University, Shanghai 200444, China (E-mail: xzhang@shu.edu.cn).}\thanks{\emph{Corresponding author: Dr. Hanzhou Wu}.}}



\maketitle

\begin{abstract}
Large Language Models (LLMs) present significant challenges to source tracing of generated code, as LLM rewriting can substantially alter the syntactic and structural characteristics of watermarked code. However, existing watermarking methods often suffer severe performance degradation under LLM-based functionality-preserving rewriting attacks. We attribute this vulnerability not merely to the intensity of rewriting, but to a fundamental distribution mismatch between the perturbations encountered during watermark training and the global, non-deterministic rewriting behaviors induced by LLMs. In order to deal with this challenge, we propose RoCoMark, a \underline{Ro}bust \underline{Co}de water\underline{Mark}ing framework that explicitly learns and adapts to LLM rewriting behaviors. RoCoMark is a multi‑round minimax game between the watermark embedder and LLM rewriter. In each round, the embedder first learns threat profiling from LLM-rewritten code by optimizing the watermark encoder through contrastive learning and knowledge distillation while mitigating catastrophic representation drift. Then the embedder improves strategies by optimizing the watermark encoder and extractor simultaneously. Finally, an adaptive attack-evolution mechanism based on Reflexion and short- and long-term memory pools is applied, enabling continuous interaction between the watermark embedder and evolving rewriting strategies. This iterative attacker-defender process progressively improves the robustness of the watermark against diverse and previously unseen LLM rewriting behaviors. Extensive experiments on four typical benchmarks, under attacks generated by five LLMs, show the effectiveness of RoCoMark. RoCoMark achieves an average detection AUROC of 74.2\% and a Rewrite BitACC of 75.3\%, while maintaining a code functionality pass rate above 99.7\%.
\end{abstract}

\begin{IEEEkeywords}
Robust code watermarking, large language models, rewriting attack, copyright protection, adversarial training.
\end{IEEEkeywords}

\section{Introduction}
\IEEEPARstart{L}{arge} language models (LLMs) have advanced code generation but raise challenges in intellectual-property protection, provenance tracking, and accountability. Prior work has explored fingerprinting and watermarking for model protection \cite{zhang2026rst,liu2024fourier}, as well as steganography for natural language \cite{yang2024stega}. These methods, however, target model behavior or natural-language content rather than source code. LLM-generated code may be modified, redistributed without authorization, or misused for academic misconduct, motivating reliable ownership and provenance verification \cite{sun2023codemark}. Source-code watermarking embeds imperceptible ownership signatures in generated code, enabling identification and provenance verification while preserving functionality.

Based on the stage at which the watermark is embedded and the underlying technical approach, existing methods can be broadly classified into two categories. The first category comprises inference-time watermarking methods, which embed signatures during the LLM decoding process by manipulating the token probability distribution over the vocabulary, without requiring additional model training or post-processing. This line of work originated from the KGW paradigm \cite{kirchenbauer2023watermark} for natural language watermarking \cite{kuditipudi2023robust,hou2024semstamp} and has since evolved into several variants adapted to the code generation setting \cite{lee2024wrote,guan2024codeip,guo2026practical,li2026efficient}. However, inference-time watermarking faces a fundamental conflict between watermark embedding and code validity. Source code is a relatively low-entropy medium governed by strict syntactic and semantic constraints; consequently, perturbing even a single token can lead to compilation failure, syntactic errors, or unintended changes in program semantics.

The second category comprises neural-based source code watermarking methods, which typically follow a post-processing paradigm. In these approaches, an LLM first generates unwatermarked code, after which a trained encoder-decoder network embeds and extracts watermark information from the code feature space. The resulting transformations must satisfy syntactic constraints while preserving the functionality of the original code. Early methods in this line of work primarily relied on predefined sets of functionality-preserving transformation rules \cite{li2022ropgen,chakraborty2022natgen,li2023protecting,xu2025clasp}. However, purely rule-based approaches are constrained by limited watermark capacity and fixed transformation patterns, making them susceptible to targeted watermark removal attacks.

To address these limitations, subsequent studies have combined rule-based transformations with learning-based networks and introduced dual-channel embedding schemes. These approaches further developed MutableAST \cite{yang2024srcmarker} as a unified abstract syntax tree (AST) based intermediate representation, together with feature-approximation techniques, to support end-to-end training and enhance watermark robustness while maintaining a high functional pass rate \cite{zhang2025robust,kim2026marking}. By performing controllable transformations under syntactic constraints, post-processing methods can achieve substantially higher functional pass rates than inference-time approaches. This advantage motivates our adoption of the post-processing paradigm in this work. For source code watermarking, functional correctness is a prerequisite for practical deployment: a watermarking method that compromises program functionality is of limited practical utility, regardless of its watermark robustness.

Despite these advances, existing neural-based source code watermarking methods remain vulnerable to a new and increasingly significant threat: functionality-preserving code rewriting by modern LLMs, such as Claude\footnote{URL: \url{https://www.anthropic.com/news/claude-sonnet-4-5}} and DeepSeek-V4-Flash \footnote{URL: \url{https://huggingface.co/deepseek-ai/DeepSeek-V4-Flash}}~\cite{suresh2024watermarking,wu2024codewmbench,cao2025robustness}. These LLMs can perform extensive rewriting while preserving the functionality of the original code, substantially exceeding the scope of the simple, predefined transformation spaces considered by RoPGen \cite{li2022ropgen} and SrcMarker \cite{yang2024srcmarker}. As a result, such rewriting attacks can significantly alter the syntactic and structural characteristics of watermarked code, potentially compromising the effectiveness of existing watermarking methods.

\begin{figure}[!t]
\centering
\includegraphics[width=\linewidth]{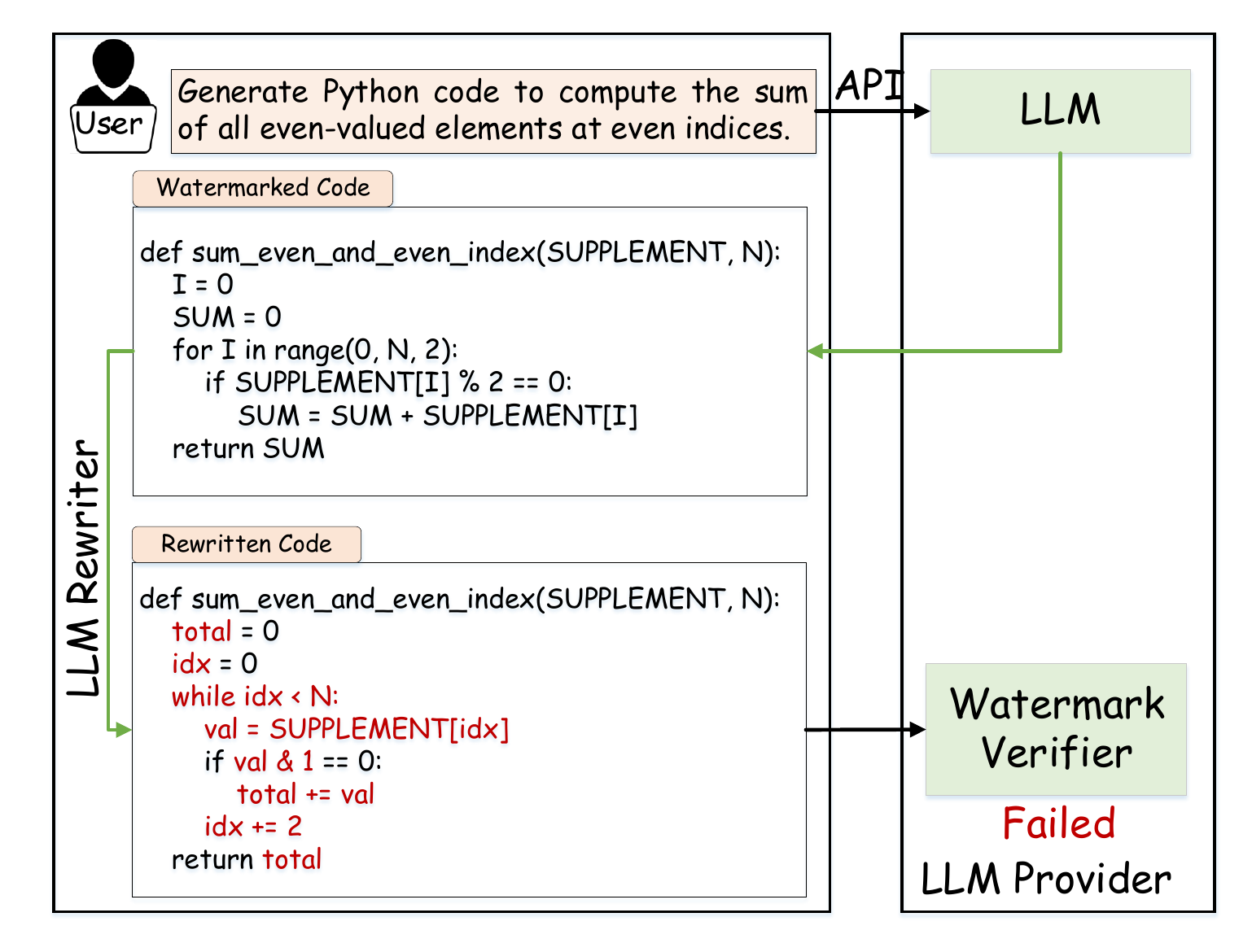}
\caption{The workflow for removing source code watermarks via LLM-powered rewriting.}
\label{fig:figure1}
\end{figure}

As shown in Fig.~\ref{fig:figure1}, a user sends a prompt to the LLM provider via an API. The LLM generates code that satisfies the requirements and embeds a watermark. After obtaining the watermarked code, the user uses any LLM to rewrite the code while preserving its functionality through operations such as variable substitution, loop structure replacement, and equivalent transformation of formulas. The LLM provider is unable to extract the watermark from the rewritten code. The fundamental reason why existing methods fail under such attacks is that their watermark encoders are not trained on adversarial samples that adequately reflect the LLM-generated rewrites. This distribution mismatch between training-time transformations and real-world LLM rewriting substantially limits the robustness of existing watermarking methods. Therefore, developing a watermarking method that can effectively withstand sophisticated LLM-powered functionality-preserving rewriting attacks remains a pressing research challenge.

\subsection{Motivation}
\begin{figure}[!t]
\centering
\includegraphics[width=0.8\linewidth]{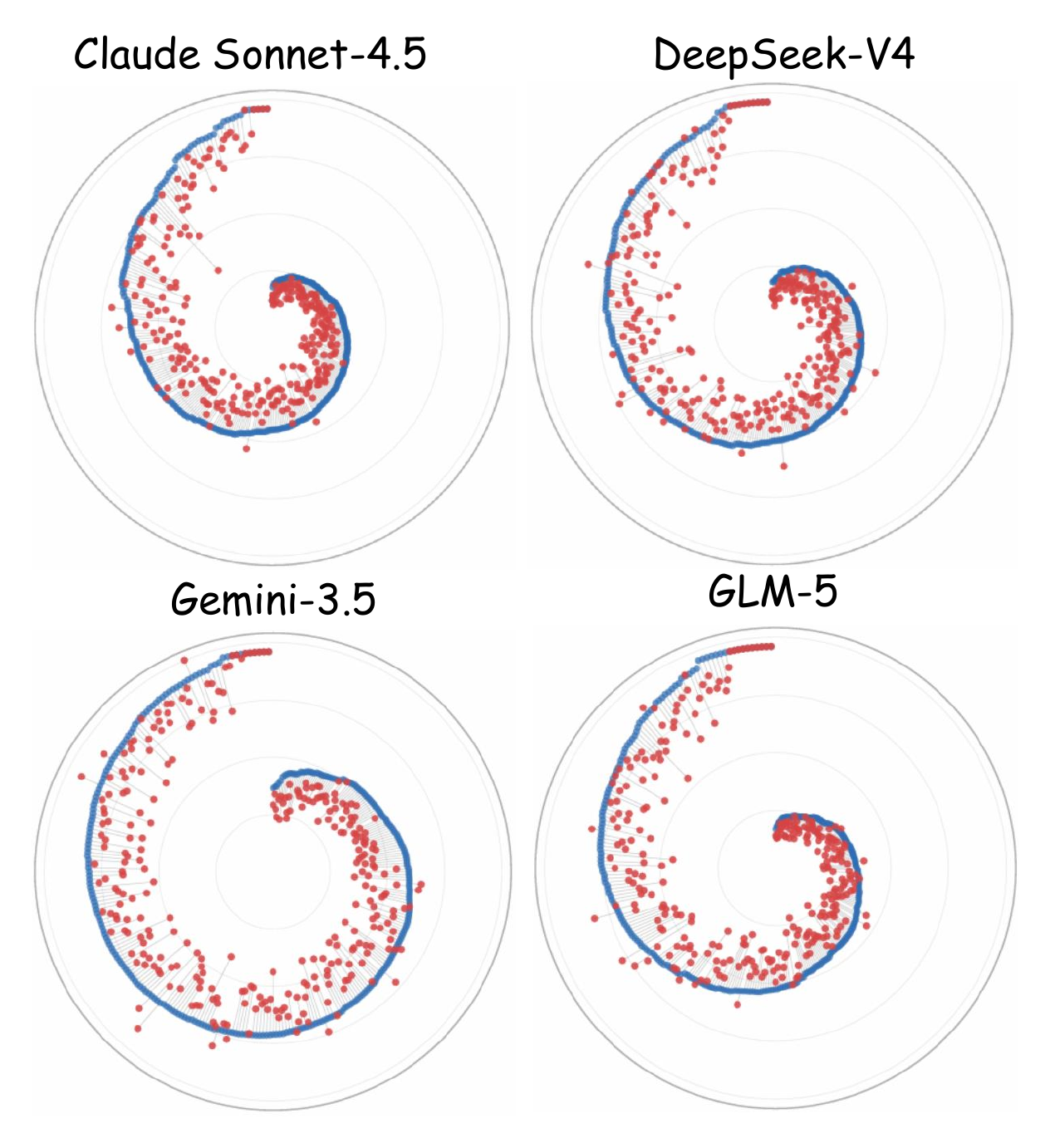}
\caption{The cosine distance distribution between the embeddings of original and rewritten code before and after CodeT5 training. The rewriting models are Claude Sonnet-4.5, DeepSeek-V4-Flash, Gemini-3.5-Flash\protect\footnotemark[4], and GLM-5\protect\footnotemark[5]. Blue and red points represent the distances before and after training, respectively, and gray lines connect paired samples. Smaller radial distances indicate greater embedding similarity.}
\label{fig:figure2}
\end{figure}

Although individual rewrites appear stochastic, our analysis of hundreds of thousands of MiMo-V2.5-Pro\footnote{URL: \url{https://mimo.mi.com/docs/en-US/news/latest/v2.5-news}}
 rewrites generated with diverse prompts revealed recurring transformation patterns. We therefore fine-tuned CodeT5 \cite{wang2021codet5} with contrastive learning on 100,000 MiMo-V2.5-Pro rewrites. As shown in Fig.~\ref{fig:figure2}, the resulting encoder substantially reduces the discrepancy between their embeddings and generalizes to unseen models, such as Claude, as well as previously unseen prompts. These results suggest that the encoder learns features that are invariant to code rewriting. However, a practical rewriting-robust watermarking system still faces three challenges. First, the space of functionality-preserving rewrites is effectively unbounded, making exhaustive enumeration infeasible. Second, aggressive fine-tuning can cause catastrophic representation drift and compromise the trained decoder. Third, the ongoing attacker-defender interaction requires continuous adaptation to newly emerging attacks.

\footnotetext[4]{URL: \url{https://deepmind.google/models/model-cards/gemini-3-5-flash/}}
\footnotetext[5]{URL: \url{https://huggingface.co/zai-org/GLM-5}}

To address these challenges, we propose RoCoMark, which models the interaction between embedder and LLM rewriter as a multi-round attacker-defender game. Within each round, we introduce a threat profiling and strategy improving mechanism that enables the embedder to first absorb the LLM rewriter's current rewrite strategy and then refine its own, while the LLM rewriter simultaneously escalates. For the watermark embedder, we integrate contrastive learning with knowledge distillation to learn rewriting-invariant representations while constraining updates to the watermark encoder. For the LLM rewriter, we introduce a Reflexion-based \cite{shinn2023reflexion} attack-strategy evolution mechanism with memory pools, which selects effective attack strategies, reflects on previous attack outcomes, and generates progressively stronger rewriting attacks. Through this iterative evolution, RoCoMark continuously enhances its robustness against increasingly sophisticated LLM-based rewriting attacks. Before the first round, we perform a cold-start warm-up by training a basic watermarking network using functionality-preserving abstract syntax tree (AST) transformations and variable-name substitutions.

\subsection{Main Contributions}
The contributions of this work are summarized as follows:

\begin{itemize}
	\item We conduct an extensive analysis of code rewriting data and leverage contrastive learning to implicitly capture the rewriting patterns induced by LLMs, thereby improving the robustness of code watermarking against functionality-preserving rewriting attacks.
	
	\item We propose a novel watermarking framework that progressively strengthens watermark robustness through an evolving attacker-defender game, in which increasingly sophisticated rewriting attacks are generated and incorporated into the training process.
	
	\item Extensive experiments on four benchmarks, including HumanEval and MBPP, under attacks from five mainstream LLMs, including GPT-5.5\footnote[6]{URL: \url{https://developers.openai.com/api/docs/models/gpt-5.5}} and DeepSeek-V4-Flash, show that RoCoMark achieves state-of-the-art performance, attaining a detection AUROC of 74.2\%, a Rewrite BitACC of 75.3\%, and a code pass rate of 99.7\%.
\end{itemize}

\section{Related Works}
\subsection{Inference-Time Watermarking}
The low entropy and rigid syntax of source code make it challenging to directly adapt natural-language watermarking techniques. KGW preselects green-list tokens and increases their sampling probabilities \cite{kirchenbauer2023watermark}, but the limited token diversity can lead to semantic distortion. SWEET restricts green-list constraints to high-entropy tokens \cite{lee2024wrote}, yet its fixed entropy threshold may not generalize well across code with varying levels of complexity. EAW dynamically adjusts the watermark strength according to local entropy \cite{shi2025entropy}, but remains limited to single-bit embedding. CodeIP enables multi-bit watermarking through a type predictor \cite{guan2024codeip}; however, prediction errors can compromise robustness under code perturbations. MCGMark encodes information through probabilistic outliers \cite{ning2024mcgmark}, making the embedded watermark susceptible to statistical smoothing. REMARK-LLM employs a learned encoder and beam search to achieve higher embedding capacity \cite{zhang2024remark}, but beam search incurs additional computational overhead and does not guarantee functional correctness under aggressive rewriting. PromptMark adopts a black-box, prompt-guided watermarking strategy \cite{fahad2026promptmark}, yet its detection performance can degrade under adaptive attacks.

Despite their differences, these methods primarily embed watermarks by manipulating token generation or selection. Such interventions inevitably alter the generated code distribution and may introduce syntactic or semantic deviations, thereby compromising functional correctness, particularly under subsequent code rewriting.

\subsection{Post-Processing Watermarking}
Post-processing watermarking embeds information through syntax- or semantics-preserving transformations applied after code generation, thereby avoiding direct interference with the generation process and offering greater flexibility in preserving functional correctness.

\textbf{Rule-Based Approaches.} Li \emph{et al}. proposed RoPGen, which learns coding styles through adversarial training with data and gradient augmentation \cite{li2022ropgen}. Although effective for style-based attribution, RoPGen relies on a fixed set of stylistic features and is not designed to withstand content-level rewriting. Chakraborty \emph{et al}. introduced NatGen, which naturalizes source code through six classes of functionality-preserving transformations \cite{chakraborty2022natgen}. While scalable, its transformation space is manually predefined and therefore cannot readily adapt to emerging or previously unseen attack patterns, leaving the defense largely static after deployment.

\textbf{Neural-Network-Enhanced Approaches.} Li \emph{et al}. combined synonym substitution with statistical testing for code watermarking \cite{li2023protecting}; however, the reliance on manually designed transformations limits its adaptability to diverse rewriting behaviors. Yang \emph{et al}. proposed SrcMarker, which integrates rule-based transformations with a neural network in an end-to-end framework \cite{yang2024srcmarker}. Although it improves upon rule-based baselines, its training relies on a fixed set of synthetic transformations and does not explicitly model rewriting induced by real-world LLMs. Zhang \emph{et al}. proposed RoSeMary, the first ML-cryptography co-design for code watermarking, combining CodeT5 with zero-knowledge verification \cite{zhang2025robust}. While providing stronger cryptographic guarantees, its security formulation does not directly address semantic-level rewriting, where the program structure and surface form may be substantially altered while preserving functionality. Javidnia \emph{et al}. introduced SWaRL, which combines reinforcement learning, compiler feedback, and Low-Rank Adaptation (LoRA)~\cite{hu2021lora} to improve watermark detection and robustness against code transformations \cite{javidnia2026swarl}. However, its training does not explicitly incorporate an LLM-based adversary, while compiler feedback primarily verifies functional correctness rather than resistance to semantic-preserving rewriting. Nie \emph{et al}. proposed MATRIX, which encodes watermarks using constrained parity-check equations with BCH codes \cite{nie2026matrix}. Although this design improves reliability against bit-level errors, it does not explicitly address the more fundamental challenge of complete structural rewriting by LLMs.

Despite their differences, existing post-processing methods largely rely on predefined transformations, fixed attack models, or static optimization objectives. More importantly, LLM-based semantic-preserving rewriting has not been explicitly incorporated as a primary adversarial threat in their design and training. Consequently, their robustness may not adequately generalize to the increasingly diverse and adaptive rewriting behaviors induced by modern LLMs.

\section{Threat Model}
Before release, the code owner embeds a source-identifying binary message into the code. After the code has been potentially modified or rewritten, the owner extracts the watermark from the suspicious code to verify its provenance \cite{lee2024wrote,guan2024codeip}. Given the original code $\mathcal{C}$ and an $L$-bit message $\mathbf{m}$, RoCoMark generates the watermarked code as
\begin{equation}
\mathcal{C}_w = \operatorname{Embed}(\mathcal{C}, \mathbf{m}).
\end{equation}
In practice, RoCoMark employs a secret key throughout the watermark embedding and extraction processes, which controls the generation and verification of the source-identifying watermark. For ease of presentation, the cryptographic key is not explicitly included in the above formulation. 

We assume that a black-box attacker possesses full access to the watermarked code $\mathcal{C}_w$, but has no prior knowledge of the original code $\mathcal{C}$ or the embedded message $\mathbf{m}$. The attacker may leverage arbitrary resources, including code LLMs, compilers, and test suites, to generate and validate rewrites. The objective of the attacker is to remove the watermark while preserving the tested behavior and public interface of $\mathcal{C}_w$. 

Mathematically, let $X = \{x_{1}, x_{2}, \dots, x_{n}\}$ represent a set of test cases, where each $x_i \in T$ specifies input arguments, execution context, and expected output $y_i$. We express $\operatorname{Exec}(\mathcal{C}, x)$ as the observable execution outcome, and let $\operatorname{Sig}(\mathcal{C})$ denote the public signature (function name, parameters, return type). A rewritten code $\mathcal{C}_{rw}$ constitutes a successful attack if
\begin{equation}
\left\{
\begin{aligned}
	&\operatorname{Exec}(\mathcal{C}_{rw}, x_i) = \operatorname{Exec}(\mathcal{C}_w, x_i) = y_i, \forall x_i\in X,\\
	&\operatorname{Sig}(\mathcal{C}_{rw}) = \operatorname{Sig}(\mathcal{C}_w),\\
	&\operatorname{WMarkDetection}(\mathcal{C}_{rw}, \mathbf{m}) < \epsilon.
\end{aligned}
\right.
\end{equation}
Here, $\operatorname{WMarkDetection}(\mathcal{C}_{rw}, \mathbf{m})$ gives a score reflecting watermark detection, which means failed watermark extraction if it is smaller than a small pre-determined threshold $\epsilon$.

To this end, the attacker feeds $\mathcal{C}_w$ and $X$ into an LLM with a prompt requesting a new implementation that maintains tested functionality and the public interface. Namely, he expects that
\begin{equation}
\left\{
\begin{aligned}
	&\mathcal{C}_{rw} = \operatorname{ReLLM}(\mathcal{C}_w, X),\\
	&\operatorname{Exec}(\mathcal{C}_{rw}, x_i) = \operatorname{Exec}(\mathcal{C}_w, x_i) = y_i, \forall x_i\in X,\\
	&\operatorname{Sig}(\mathcal{C}_{rw}) = \operatorname{Sig}(\mathcal{C}_w).
\end{aligned}
\right.
\end{equation}
Here, $\operatorname{ReLLM}$ corresponds to the LLM used for rewriting. Due to the stochastic nature of LLM generation, repeated queries can produce structurally diverse implementations. LLM-based rewriting reconstructs code by reasoning about program semantics, data dependencies, and control logic. Consequently, it constitutes a composite attack that can simultaneously alter identifiers, expressions, data flow, and control flow, making it substantially more comprehensive than attacks targeting a single type of code transformation \cite{li2022ropgen,chakraborty2022natgen,yang2024srcmarker,zhang2025robust}.

\textbf{Design Goals.} Based on the above threat model, RoCoMark is designed to satisfy the following four objectives:

\begin{itemize}
	\item \emph{Effectiveness:} The watermark should be accurately embedded into the original code and reliably recovered from the watermarked code, while preserving the intended program functionality.
	\item \emph{Imperceptibility:} The watermarked code should remain natural and semantically consistent with the original code, with all transformations satisfying the prescribed AST constraints and employing natural-looking identifiers.
	\item \emph{Robustness:} The embedded watermark should remain detectable after realistic LLM-based rewriting attacks, rather than being evaluated only against fixed, rule-based transformations.
	\item \emph{Capacity:} The watermark should provide sufficient payload capacity to encode source-identifying information and distinguish among different code owners.
\end{itemize}

\section{Code Watermarking via RoCoMark}
\subsection{System Overview}
\begin{figure*}[!htbp]
\centering
\includegraphics[width=0.7\textwidth]{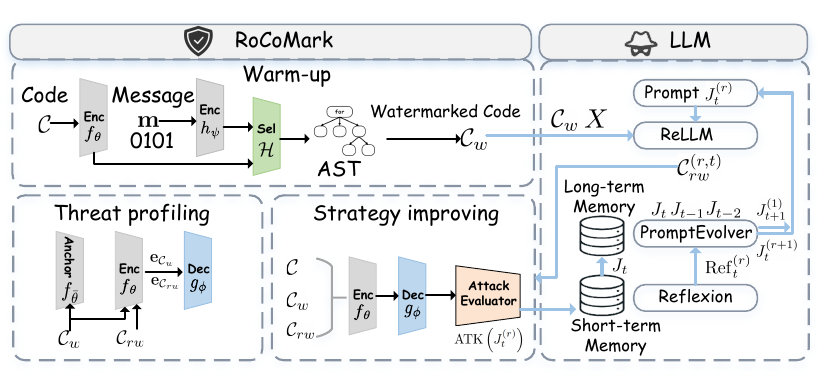}
\caption{Overview of RoCoMark. The defender comprises threat profiling, strategy improving, warm-up to avoid cold start. The attacker introduces a Reflexion-based attack-strategy evolution mechanism with memory pools.}
\label{fig:RoCoMark}
\end{figure*}

RoCoMark achieves these goals through a multi-round attacker-defender game, as illustrated in Fig.~\ref{fig:RoCoMark}. At each round, the LLM attacker selects effective strategies, reflects on previous attempts, and generates increasingly powerful rewriting prompts. The resulting watermark-disruptive rewrites are then fed into robustness training, enabling the watermark model to continuously adapt throughout the game. The defender is built upon two complementary mechanisms. \textit{Threat profiling} employs an LLM to rewrite the watermarked code and optimizes the code encoder via contrastive learning, thereby implicitly capturing the distribution of rewriting patterns. \textit{Strategy improving} strengthens the watermark itself through syntax-and-variable dual-channel embedding, making it more resilient to the observed attacks. Before entering the adversarial loop, a \textit{Warm-up} phase initializes the basic embedding and extraction capabilities on public datasets using the same dual-channel approach, effectively mitigating the cold-start problem.

\subsubsection{Watermark Embedding}
RoCoMark uses pre-trained CodeT5 as the code encoder. Let $f_\theta$ denote the code encoder parameterized by $\theta$, and let $h_\psi$ denote the message encoder parameterized by $\psi$. The original code representation and the message representation are respectively given by $\mathbf{e}_\mathcal{C} = f_\theta(\mathcal{C})$ and $\mathbf{v} = h_\psi(\mathbf{m})$. The syntax-selection head $\mathcal{H}_{\mathrm{syn}}$ and the variable-selection head $\mathcal{H}_{\mathrm{var}}$ fuse $\mathbf{e}_\mathcal{C}$ and $\mathbf{v}$ and produce the following outputs, respectively:
\begin{equation}
\left\{
\begin{aligned}
    &\mathbf{p}_{\mathrm{syn}} = \operatorname{MaskedSoftmax}\left(\mathcal{H}_{\mathrm{syn}}(\mathbf{e}_\mathcal{C},\mathbf{v}); \mathcal{A}_{\mathrm{syn}}(\mathcal{C})\right),\\
    &\mathbf{p}_{\mathrm{var}} = \operatorname{MaskedSoftmax}\left(\mathcal{H}_{\mathrm{var}}(\mathbf{e}_\mathcal{C},\mathbf{v}); \mathcal{A}_{\mathrm{var}}(\mathcal{C})\right).
\end{aligned}
\right.
\end{equation}
Here, $\mathcal{H}_{\mathrm{syn}}$ and $\mathcal{H}_{\mathrm{var}}$ denote the syntax-selection and variable-selection heads, respectively, while $\mathcal{A}_{\mathrm{syn}}(\mathcal{C})$ and $\mathcal{A}_{\mathrm{var}}(\mathcal{C})$ represent the corresponding sets of valid syntax actions and variable-name candidates for the current code. During training, hard Gumbel-Softmax with temperature $\tau_g$ is used to approximate discrete selection; at inference, the highest-probability action is selected directly. The selected actions and the resulting watermarked code are then given by
\begin{equation}
\left\{
\begin{aligned}
&a_{\mathrm{syn}} = \operatorname{arg~max}_{s\in \mathcal{A}_\mathrm{syn}(\mathcal{C})} \mathbf{p}_{\mathrm{syn}},\\
&a_{\mathrm{var}} = \operatorname{arg~max}_{v\in \mathcal{A}_\mathrm{var}(\mathcal{C})} \mathbf{p}_{\mathrm{var}},\\
&\mathcal{C}_w = \operatorname{Transform}(\mathcal{C}, a_{\mathrm{syn}}, a_{\mathrm{var}}).
\end{aligned}
\right.
\end{equation}
Here, $\operatorname{Transform}$ is a deterministic executor that applies the selected syntax action and variable-name substitution rule to the AST and regenerates the code. While syntactic validity is jointly enforced by the validity mask and the rule executor, functional preservation is ensured through the deliberate design of functionality-preserving transformations and subsequent test-based validation, rather than being presumed from a free-form text generator.

\subsubsection{Watermark Extraction}
The watermark extraction module consists of a shared CodeT5 encoder $f_\theta$ and a lightweight watermark decoder $g_\phi$. For a suspected code $\mathcal{C}_s$, the per-bit watermark logits are defined as $\mathbf{s}^{\mathcal{C}_s}_{\mathrm{pred}} = g_\phi(f_\theta(\mathcal{C}_s)) \in \mathbb{R}^L$, where $g_\phi$ denotes the watermark decoder parameterized by $\phi$ and $L$ is the watermark message length. The corresponding hard-decision results are given by
\begin{equation}
\hat{m}_{\mathcal{C}_s,i} = I\!\left[ \sigma\!\left( {s}^{\mathcal{C}_s}_{\mathrm{pred},i} \right) > \delta \right], i = 1, 2, \dots, L.
\end{equation}
where $\delta$ is the per-bit decision threshold (specified in the experimental configuration), and $I[\cdot]$ is the indicator function. The predicted watermarks for the watermarked code and the rewritten code are denoted by $\hat{\mathbf{m}}_{\mathcal{C}_w}$ and $\hat{\mathbf{m}}_{\mathcal{C}_{rw}}$, respectively.

\subsubsection{Watermarking Strength}\label{sec:watermarking-strength}
RoCoMark quantifies watermarking strength via a $z$-score under the null hypothesis of random message matching. Specifically, we define
\begin{equation}
\mathcal{K}(\mathcal{C}_s; \mathbf{m}) = \sum_{i=1}^{L} I\!\left[ \hat{{m}}_{\mathcal{C}_s, i} = {m}_i \right],\label{eq:k}
\end{equation}
where ${m}_i$ denotes the $i$-th bit of the target watermark message $\mathbf{m}$, and $\mathcal{K}(\mathcal{C}_s; \mathbf{m})$ counts the number of matching bits between the predicted message for the suspected code $\mathcal{C}_s$ and the target message $\mathbf{m}$. Under the random-matching assumption, let $\bar{p}$ denote the per-bit matching probability. The watermark detection score for $\mathcal{C}_s$ is then
\begin{equation}
\mathrm{WMarkDetection}(\mathcal{C}_s, \mathbf{m}) = \frac{\mathcal{K}(\mathcal{C}_s; \mathbf{m}) - \mu}{\sigma_k},\label{eq:WMarkDetection}
\end{equation}
where the mean and variance of the number of matching bits are $\mu = L\bar{p}$ and $\sigma_k^2 = L\bar{p}(1 - \bar{p})$, respectively. When $\mathrm{WMarkDetection}(\mathcal{C}_s, \mathbf{m}) \geq \epsilon$, the suspected code is classified as containing the target watermark.

\subsection{Iterative Adversarial Co-Evolution}
RoCoMark optimizes the watermarking model within an iterative adversarial co-evolution framework, comprising three components. The \textit{Warm-up} module establishes basic watermarking strategies that preserve code functionality while ensuring detectability. The \textit{Threat profiling} module learns representations that remain stable between watermarked and LLM-rewritten code. The \textit{Strategy improving} module jointly enhances robustness against high-intensity rewriting and suppresses false positives on unwatermarked code. To drive this co-evolution, the LLM attacker employs a Reflexion-based attack-strategy evolution mechanism equipped with short- and long-term memory pools.

\subsubsection{Warm-up}
To mitigate the cold-start problem of the defender in the adversarial iteration process, we pre-train a base watermark embedding model using variable substitution and AST transformations. However, these code transformations are non-differentiable, which prevents the watermark decoding loss from being directly backpropagated to the encoder. Motivated by SrcMarker \cite{yang2024srcmarker}, we introduce $\widetilde{\mathbf{e}}_{\mathcal{C}_w}$, the approximated representation produced by the feature approximator, i.e.,
\begin{equation}
\mathbf{e}_{\mathcal{C}_w}=f_\theta(\mathcal{C}_w),\quad
\widetilde{\mathbf{e}}_{\mathcal{C}_w}
=q_\omega\left(f_\theta(\mathcal{C}),\mathbf{p}_{\mathrm{syn}},\mathbf{p}_{\mathrm{var}}\right).
\end{equation}
Here, $\mathbf{e}_{\mathcal{C}_w}$ denotes the CodeT5-encoded representation of the watermarked code, while $\widetilde{\mathbf{e}}_{\mathcal{C}_w}$ is its approximation produced by the feature approximator $q_\omega$. During training, Gaussian noise is injected into the decoding branch associated with $\widetilde{\mathbf{e}}_{\mathcal{C}_w}$.
\begin{equation}
\widetilde{\mathbf{e}}_{\mathcal{C}_w}^{\varepsilon}=\widetilde{\mathbf{e}}_{\mathcal{C}_w}+\varepsilon,~\text{where}~\varepsilon\sim \mathcal{N}\left(0,\sigma_{\varepsilon}^2 I_d\right).
\end{equation}
where $I_d$ is the $d$-dimensional identity matrix. 

Warm-up uses three loss terms. The first is the approximated decoding loss:
$L_{\mathrm{approx}}^{(1)} =\operatorname{BCELogits}\!\left(g_\phi\bigl(\widetilde{\mathbf{e}}_{\mathcal{C}_w}^{\varepsilon}\bigr), \mathbf{m}\right)$, where $\operatorname{BCELogits}$ denotes the BCEWithLogits loss averaged over the $L$ watermark bits. This term back-propagates gradients through the continuous approximated representation to the message encoder, transformation selector, and feature approximator.
The second term is the actual decoding loss, which can be expressed as
$L_{\mathrm{actual}}^{(1)}=\operatorname{BCELogits}\!\left(g_\phi(\mathbf{e}_{\mathcal{C}_w}), \mathbf{m}\right)$, enabling the decoder to recover the message from code produced by executing variable substitutions and AST transformations. The third term is the feature approximation loss is $L_{\mathrm{feat}}^{(1)} =\operatorname{MSE}\!\left(\widetilde{\mathbf{e}}_{\mathcal{C}_w},\mathbf{e}_{\mathcal{C}_w}\right)$, which aligns the approximated representation $\widetilde{\mathbf{e}}_{\mathcal{C}_w}$ (before noise injection) with the actual representation $\mathbf{e}_{\mathcal{C}_w}$. The complete objective of warm-up is therefore given by
\begin{equation}
L_{\mathrm{WU}}
=
\lambda_{\mathrm{approx}}^{(1)}L_{\mathrm{approx}}^{(1)}
+
\lambda_{\mathrm{actual}}^{(1)}L_{\mathrm{actual}}^{(1)}
+
\lambda_{\mathrm{feat}}^{(1)}L_{\mathrm{feat}}^{(1)},
\end{equation}
where $\lambda_{\mathrm{approx}}^{(1)}$, $\lambda_{\mathrm{actual}}^{(1)}$, and $\lambda_{\mathrm{feat}}^{(1)}$ are used for loss balance. 

\subsubsection{Threat profiling}
Let $f_{\theta}$ be the trainable code encoder, and $f_{\bar{\theta}}$ denote the parameter-frozen anchor encoder copied from warm-up. Three types of code representations are defined as follows:
\begin{equation}
\left\{
\begin{aligned}
    &\mathbf{e}_{\mathcal{C}_w} = f_\theta(\mathcal{C}_w),\\
    &\mathbf{e}_{\mathcal{C}_{rw}} = f_\theta(\mathcal{C}_{rw}),\\
    &\bar{\mathbf{e}}_{\mathcal{C}_w} = \operatorname{detach}\!\left(f_{\bar{\theta}}(\mathcal{C}_w)\right),
\end{aligned}
\right.
\end{equation}
where $\mathbf{e}_{\mathcal{C}_w}$ and $\mathbf{e}_{\mathcal{C}_{rw}}$ are the representations of the watermarked code and the LLM-rewritten code extracted by the target encoder, respectively. $\bar{\mathbf{e}}_{\mathcal{C}_w}$ denotes the warm-up representation produced by the anchor encoder. The operator $\operatorname{detach}(\cdot)$ stops gradient computation, preventing gradients from flowing back to $f_{\bar{\theta}}$ through the anchor representation. The frozen watermark decoder then outputs
\begin{equation}
\mathbf{s}^{\mathcal{C}_w}_{\mathrm{pred}}=g_\phi(\mathbf{e}_{\mathcal{C}_w}),\mathbf{s}^{\mathcal{C}_{rw}}_{\mathrm{pred}}=g_\phi(\mathbf{e}_{\mathcal{C}_{rw}}).
\end{equation}
The decoding losses are defined as
\begin{equation}
\left\{
\begin{aligned}
    &L_{\mathrm{w}}^{(2)} = \operatorname{BCELogits}\!\left(\mathbf{s}^{\mathcal{C}_w}_{\mathrm{pred}}, \mathbf{m}\right),\\
    &L_{\mathrm{rw}}^{(2)} = \operatorname{BCELogits}\!\left(\mathbf{s}^{\mathcal{C}_{rw}}_{\mathrm{pred}}, \mathbf{m}\right),
\end{aligned}
\right.
\end{equation}
where $\mathbf{s}^{\mathcal{C}_w}_{\mathrm{pred}}$ and $\mathbf{s}^{\mathcal{C}_{rw}}_{\mathrm{pred}}$ are the $L$-dimensional watermark logits corresponding to the watermarked code and LLM-rewritten code, respectively, and $\mathbf{m}$ is the corresponding target message. The watermark recovery loss is expressed as
\begin{equation}
L_{\mathrm{wm}}^{(2)}
=
\lambda_{\mathrm{w}}^{(2)}L_{\mathrm{w}}^{(2)}
+
\lambda_{\mathrm{rw}}^{(2)}L_{\mathrm{rw}}^{(2)}.
\end{equation}

To prevent multiple rewritten variants of the same watermarked code from being erroneously treated as negative samples, RoCoMark adopts bidirectional grouped contrastive learning. The representation normalization is redefined as:
\begin{equation}
\left\{
\begin{aligned}
    &\hat{\mathbf{e}}_{\mathcal{C}_{w}} = \frac{\mathbf{e}_{\mathcal{C}_{w}}}{\max\left(\left\lVert \mathbf{e}_{\mathcal{C}_{w}}\right\rVert_2,\varepsilon_n\right)},\\
    &\hat{\mathbf{e}}_{\mathcal{C}_{rw}} = \frac{\mathbf{e}_{\mathcal{C}_{rw}}}{\max\left(\left\lVert \mathbf{e}_{\mathcal{C}_{rw}}\right\rVert_2,\varepsilon_n\right)}.
\end{aligned}
\right.
\end{equation}

Positive pairs are formed by a watermarked code and its corresponding LLM-rewritten variant that originate from the same original code. Let $\hat{\mathbf{e}}_{\mathcal{C}_w}^+$ denote the normalized representation of a watermarked code, and $\hat{\mathbf{e}}_{\mathcal{C}_{rw}}^+$ denote the normalized representation of its paired rewritten version. All other rewritten representations constitute the negative samples.

The bidirectional contrastive losses are defined as
\begin{equation}
\left\{
\begin{aligned}
&L_{\mathrm{w}\rightarrow\mathrm{rw}} = -\log \frac{ \exp\bigl( \hat{\mathbf{e}}_{\mathcal{C}_{w}}^\top \hat{\mathbf{e}}_{\mathcal{C}_{rw}}^+ / \tau_c \bigr) }{ \displaystyle\sum_{\hat{\mathbf{e}}_{\mathcal{C}_{rw}}^+} \exp\bigl(\hat{\mathbf{e}}_{\mathcal{C}_{w}}^\top \hat{\mathbf{e}}_{\mathcal{C}_{rw}}^+ / \tau_c \bigr) },\\
&L_{\mathrm{rw}\rightarrow \mathrm{w}} = -\log \frac{ \exp\bigl( \hat{\mathbf{e}}_{\mathcal{C}_{rw}}^\top \hat{\mathbf{e}}_{\mathcal{C}_{w}}^+ / \tau_c \bigr) }{ \displaystyle\sum_{\hat{\mathbf{e}}_{\mathcal{C}_{w}}^+} \exp\bigl( \hat{\mathbf{e}}_{\mathcal{C}_{rw}}^\top \hat{\mathbf{e}}_{\mathcal{C}_{w}}^+ / \tau_c \bigr) }.
\end{aligned}
\right.
\end{equation}
where $\tau_c$ is the temperature parameter, and the expectation is computed uniformly over all positive pairs. The grouped contrastive loss is then defined as
\begin{equation}
L_{\mathrm{cont}}^{(2)}
=
\frac{1}{2}
\left(
L_{\mathrm{w}\rightarrow\mathrm{rw}}
+
L_{\mathrm{rw}\rightarrow \mathrm{w}}
\right).
\end{equation}

To constrain the target encoder from drifting away from the warm-up representation space, RoCoMark introduces a representation distillation constraint. The full set of auxiliary losses comprises four terms: the warm-up representation distillation loss $L_{\mathrm{distill}}^{(2)}$, the adversarial-anchor loss $L_{\mathrm{rw\text{-}anchor}}^{(2)}$, the cosine alignment loss $L_{\mathrm{align}}^{(2)}$, and the decoding-difficulty consistency loss $L_{\mathrm{cons}}^{(2)}$:
\begin{equation}
\left\{
\begin{aligned}
    &L_{\mathrm{distill}}^{(2)} = \operatorname{MSE}\!\left(\mathbf{e}_{\mathcal{C}_{w}},\bar{\mathbf{e}}_{\mathcal{C}_{w}}\right),\\
    &L_{\mathrm{rw\text{-}anchor}}^{(2)} = \operatorname{MSE}\!\left(\mathbf{e}_{\mathcal{C}_{rw}},\bar{\mathbf{e}}_{\mathcal{C}_{w}}\right),\\
    &L_{\mathrm{align}}^{(2)} = 1 - \cos\!\left(\mathbf{e}_{\mathcal{C}_{w}},\mathbf{e}_{\mathcal{C}_{rw}}\right),\\
    &L_{\mathrm{cons}}^{(2)} = \left|L_{\mathrm{w}}^{(2)} - L_{\mathrm{rw}}^{(2)}\right|.
\end{aligned}
\right.
\end{equation}
Specifically, the decoding-difficulty consistency loss is computed by comparing the BCE values of the two paths with respect to the target message, rather than requiring the two hard predictions to match exactly. The complete threat profiling training objective is then formulated as
\begin{equation}
\begin{aligned}
L&_{\mathrm{TP}}
=
\lambda_{\mathrm{wm}}^{(2)}L_{\mathrm{wm}}^{(2)}
+
\lambda_{\mathrm{cont}}^{(2)}L_{\mathrm{cont}}^{(2)}
+
\lambda_{\mathrm{distill}}^{(2)}L_{\mathrm{distill}}^{(2)}
\\
&+
\lambda_{\mathrm{rw\text{-}anchor}}^{(2)}
L_{\mathrm{rw\text{-}anchor}}^{(2)}
+
\lambda_{\mathrm{align}}^{(2)}L_{\mathrm{align}}^{(2)}
+
\lambda_{\mathrm{cons}}^{(2)}L_{\mathrm{cons}}^{(2)}.
\end{aligned}
\end{equation}

The threat profiling module updates only the target CodeT5 encoder $f_\theta$. The anchor encoder $f_{\bar\theta}$, watermark decoder $g_\phi$, message encoder, transformation selector, and feature approximator are all kept frozen.

\subsubsection{Strategy improving}
The rewriting intensity induced by different prompts varies substantially. Simple prompts typically modify only variable names and localized expressions, whereas strong prompts may simultaneously alter control flow, loop structures, intermediate computations, and return logic. To enhance adaptability to high-intensity rewriting, RoCoMark constructs a training set from the generated valid code and resumes training.

During strategy improvement, RoCoMark regenerates $\mathcal{C}_w$ for the current model and samples $\mathcal{C}_{rw}$ from the offline pool that corresponds to the same original data. A hard Gumbel-Softmax with temperature $\tau_g$ is used to select watermark carriers during training, while $\operatorname*{argmax}$ is employed during evaluation. The three types of decoding logits are defined as
\begin{equation*}
\mathbf{s}^{\mathcal{C}}_{\mathrm{pred}} = g_\phi\!\left(f_\theta(\mathcal{C})\right),\mathbf{s}^{\mathcal{C}_w}_{\mathrm{pred}} = g_\phi\!\left(f_\theta(\mathcal{C}_w)\right),\mathbf{s}^{\mathcal{C}_{rw}}_{\mathrm{pred}} = g_\phi\!\left(f_\theta(\mathcal{C}_{rw})\right).
\end{equation*}
The decoding loss on watermarked code is given by
\begin{equation}
L_{\mathrm{w}}^{(3)}
=
\operatorname{BCELogits}\left(
\mathbf{s}^{\mathcal{C}_{w}}_{\mathrm{pred}},\mathbf{m}
\right).
\end{equation}
Let $\rho$ indicate whether the rewrite passes the functionality check, where $\rho = 1$ if it passes and $\rho = 0$ otherwise. The adversarial-loss weight is then defined as
\begin{equation}
v = I\!\left[\,\rho = 1 \;\wedge\; \mathcal{C}_{rw} \neq \emptyset \;\wedge\; \operatorname{strip}(\mathcal{C}_{rw}) \neq \operatorname{strip}(\mathcal{C}_w)\,\right],
\end{equation}
where $\operatorname{strip}(\cdot)$ removes leading and trailing whitespace from the code. The adversarial decoding loss for a single valid rewritten sample is expressed as
\begin{equation}
L^{(3)}_{\text{rw}} = v \cdot \operatorname{BCELogits}\!\left(\mathbf{s}^{\mathcal{C}_{rw}}_{\mathrm{pred}}, \mathbf{m}\right).
\end{equation}
When the sample is invalid (i.e., $v = 0$), the loss is zero and contributes nothing to the overall objective.

As described earlier, $\widetilde{\mathbf{e}}_{\mathcal{C}_w}$ denotes the approximated representation produced by the current watermark selection, and $\mathbf{e}_{\mathcal{C}_w} = f_\theta(\mathcal{C}_w)$ is the actual watermarked-code representation. The representation proxy loss is $L_{\mathrm{feat}}^{(3)}=\operatorname{MSE}\left(\widetilde{\mathbf{e}}_{\mathcal{C}_w},\mathbf{e}_{\mathcal{C}_w}\right)$. The discrete validation failure rate is defined as $L_{\mathrm{guard}}^{(3)}=1-\rho$.

To construct a differentiable detection margin constraint, RoCoMark adopts a soft $z$-score formulation. Namely, given $\mathcal{C}$ or $\mathcal{C}_{rw}$, the soft matching probability for the $k$-th bit is
\begin{equation}
\left\{
\begin{aligned}
    &q_{k}^{\mathcal{C}} = m_k\,\sigma\!\left(\mathbf{s}^{\mathcal{C}}_{\mathrm{pred}}\right) + (1-m_k)\!\left[1-\sigma\!\left(\mathbf{s}^{\mathcal{C}}_{\mathrm{pred}}\right)\right],\\
    &q_{k}^{\mathcal{C}_{rw}} = m_k\,\sigma\!\left(\mathbf{s}^{\mathcal{C}_{rw}}_{\mathrm{pred}}\right) + (1-m_k)\!\left[1-\sigma\!\left(\mathbf{s}^{\mathcal{C}_{rw}}_{\mathrm{pred}}\right)\right].
\end{aligned}
\right.
\end{equation}

This expression returns the predicted probability of the target bit being $1$ or $0$, depending on the ground truth. The corresponding soft detection score is determined by
\begin{equation}
\left\{
\begin{aligned}
    &z_{\mathrm{soft}}^{\mathcal{C}} = \frac{\displaystyle\sum_{k=1}^{L}q_{k}^{\mathcal{C}}-L\bar{p}}{\sqrt{L\bar{p}\left(1-\bar{p}\right)}},\\
    &z_{\mathrm{soft}}^{\mathcal{C}_{rw}} = \frac{\displaystyle\sum_{k=1}^{L}q_{k}^{\mathcal{C}_{rw}}-L\bar{p}}{\sqrt{L\bar{p}\left(1-\bar{p}\right)}}.
\end{aligned}
\right.
\end{equation}
where $\bar{p}$ denotes the per-bit matching probability. The detection margin loss is
\begin{equation}
L_{\mathrm{margin}}^{(3)}
=
v
\left[
\zeta
-
z_{\mathrm{soft}}^{\mathcal{C}_{rw}}
+
z_{\mathrm{soft}}^{\mathcal{C}}
\right]_+.
\end{equation}
Here, $[x]_+=\max(0,x)$. This loss requires the soft detection score of valid rewritten code to be at least $\zeta$ higher than that of the corresponding original code. The original code contrastive loss is $L_{\mathrm{neg}}^{(3)}=\operatorname{BCELogits}\left(\mathbf{s}^{\mathcal{C}}_{\mathrm{pred}},1-\mathbf{m}\right)$, where $1-\mathbf{m}$ is the bitwise complement of the target message $\mathbf{m}$. This term pushes the predictions for original unwatermarked code away from the target watermark, thereby reducing false positives caused by random matching. Thus, the final loss function is
\begin{equation}
\begin{aligned}
L_{\mathrm{SI}}
={}&
\lambda_{\mathrm{w}}^{(3)}
L_{\mathrm{w}}^{(3)}
+
\lambda_{\mathrm{rw}}^{(3)}
L_{\mathrm{rw}}^{(3)}
+
\lambda_{\mathrm{feat}}^{(3)}
L_{\mathrm{feat}}^{(3)}
\\
&+
\lambda_{\mathrm{guard}}^{(3)}
L_{\mathrm{guard}}^{(3)}
+
\lambda_{\mathrm{margin}}^{(3)}
L_{\mathrm{margin}}^{(3)}
+
\lambda_{\mathrm{neg}}^{(3)}
L_{\mathrm{neg}}^{(3)}.
\end{aligned}
\end{equation}

In summary, this module jointly updates the shared CodeT5 encoder, watermark encoder, watermark decoder, syntax and variable transformation selectors, and feature approximator. And, the warm-up anchor encoder remains frozen throughout.

\subsubsection{LLM-Based Prompt Evolution}
Fixed prompts generally only cover predefined rewriting patterns and cannot adapt to continuously evolving LLM-based attacks. To overcome this limitation, RoCoMark introduces a Reflexion-based prompt evolution mechanism \cite{shinn2023reflexion} within the threat profiling and strategy improvement phases. Unlike conventional genetic algorithms relying on crossover, mutation, and population-based optimization \cite{fernando2023promptbreeder}, here evolution refers to iterative refinement of prompt content by an LLM driven by prior attack outcomes.

Let $t$ denote the outer adversarial round, and let $r$ denote the inner iteration index within round $t$. At each round, the LLM attacker performs multiple inner iterations to search for a prompt that effectively reduces the watermark detection score. Let $J_t^{(r)}$ be the $r$-th code rewriting prompt at round $t$. Initially, RoCoMark uses hand-crafted seed prompts to generate the first batch of rewritten outputs. For a watermarked code $\mathcal{C}_w$ and test cases $X$, the rewriting process is expressed as
\begin{equation*}
\mathcal{C}_{rw}^{(r,t)}
=
\operatorname{ReLLM}\left(\mathcal{C}_{w},X,J_t^{(r)}\right).
\end{equation*}
where $\operatorname{ReLLM}$ is the previously defined LLM-based code rewriting operation. The generated code must pass syntax, interface-consistency, and functionality-consistency checks. The corresponding validity indicator is denoted by $\rho^{(r,t)}$, where $\rho^{(r,t)} = 1$ indicates a valid rewrite and $\rho^{(r,t)} = 0$ otherwise. The attack effectiveness of prompt $J_t^{(r)}$ is
\begin{multline}
\operatorname{Atk}\left(J_t^{(r)}\right)
= \rho^{(r,t)}
\Bigl[
\mathrm{WMarkDetection}(\mathcal{C}_w, \mathbf{m}) \\
- \mathrm{WMarkDetection}(\mathcal{C}_{rw}^{(r,t)}, \mathbf{m})
\Bigr]_{+}.
\end{multline}
A higher score indicates that the prompt produces valid rewrites that more strongly disrupt the watermark. This score is used solely to construct adversarial training samples and is not disclosed to external attackers.

Within each round $t$, the LLM maintains a short-term memory that stores all attempted prompts $J_t^{(r)}$ and their corresponding attack outcomes from the current inner iterations. Based on this short-term memory, the LLM generates a reflection summary $\mathrm{Ref}_t^{(r)}$ after each inner iteration, capturing why certain prompts succeeded or failed. The reflection guides the PromptEvolver to generate the next candidate prompt $J_t^{(r+1)} = \operatorname{PromptEvolver}\Bigl( J_t^{(r)},\; \mathrm{Ref}_t^{(r)} \Bigr)$. This inner loop repeats until a prompt achieves a sufficiently high attack effectiveness (i.e., successfully defeats the current watermark model). Once such a prompt is found, it is stored together with its corresponding reflection into the long-term memory. The long-term memory accumulates the strongest attacking prompts and their associated reflections across different rounds. The defender (watermark model) then undergoes threat profiling and strategy improvement specifically against this successful prompt. After the defender is updated to resist it, the adversarial process proceeds to the next round $t+1$.

At the beginning of round $t+1$, the LLM retrieves the three most recent entries from the long-term memory and uses them as initial seeds. The PromptEvolver operation then generates the first prompt for the new round, i.e., $J_{t+1}^{(1)} = \operatorname{PromptEvolver}\left(J_{t}, J_{t-1}, J_{t-2} \right)$, where $J_{t}, J_{t-1}, J_{t-2}$ are the three most recently stored prompts from the long-term memory. The inner iterations then resume, continuing the co-evolutionary cycle.

\subsection{Watermark Verification}
After training, given a suspected code $\mathcal{C}_{s}$ to be verified, the authorized verifier first computes its watermark decoding logits $\mathbf{s}^{\mathcal{C}_{s}}_{\mathrm{pred}}=g_\phi\left(f_\theta\left(\mathcal{C}_{s}\right)\right)$. The predicted $i$-th message bit is
$\hat{{m}}_{{\mathcal{C}_{s}},i}=I\left[\sigma\left({s}^{\mathcal{C}_{s}}_{\mathrm{pred},i}\right)>\delta\right]$. The system computes the hard detection score $\mathrm{WMarkDetection}(\mathcal{C}_s, \mathbf{m})$ from the predicted message $\hat{\mathbf{m}}_{\mathcal{C}_{s}}$ and the target message $\mathbf{m}$ held by the code owner. The final decision is the same as Eq.~(\ref{eq:WMarkDetection}). Note that, after training, only the code encoder, watermark decoder, and target watermark message held by the code owner need to be retained at deployment. The feature approximator, anchor encoder, and LLM rewriting module are used only during training and do not participate in watermark verification at deployment.

\section{Experimental Results and Analysis}
\subsection{Setup}
\subsubsection{Hyperparameter Settings}
We train the model in three stages, each with its own set of hyperparameters. In the \textit{Warm-up} stage, we empirically set $\lambda_{\mathrm{approx}}^{(1)}= \lambda_{\mathrm{actual}}^{(1)}=4\lambda_{\mathrm{feat}}^{(1)}=1.0$. In the \textit{Threat profiling} stage, we set $\lambda_{\mathrm{w}}^{(2)}=1.0$, $\lambda_{\mathrm{rw}}^{(2)}=1.5$, $\lambda_{\mathrm{cont}}^{(2)}=0.1$, $\lambda_{\mathrm{wm}}^{(2)}=1.0$, $\lambda_{\mathrm{distill}}^{(2)}=0.5$, $\lambda_{\mathrm{rw-anchor}}^{(2)}=1.0$, $\lambda_{\mathrm{align}}^{(2)}=0.2$, and $\lambda_{\mathrm{cons}}^{(2)}=0.1$. In the \textit{Strategy improving} stage, we set $\lambda_{\mathrm{w}}^{(3)}=1.0$, $\lambda_{\mathrm{rw}}^{(3)}=2.0$, $\lambda_{\mathrm{feat}}^{(3)}=0.25$, $\lambda_{\mathrm{guard}}^{(3)}=0.05$, $\lambda_{\mathrm{margin}}^{(3)}=0.3$, and $\lambda_{\mathrm{neg}}^{(3)}=0.05$. The global hyperparameters are kept fixed across all three stages, with $\epsilon=1.645$, $\sigma_{\varepsilon}=0.02$, $\varepsilon_n=10^{-12}$, $L=4$, $\tau_g=0.5$, $\tau_c=0.07$, $\delta=0.5$, $\bar{p}=0.5$, and $\zeta=1.0$. Although further hyperparameter tuning may yield performance gains, an exhaustive search would be labor-intensive and is not the primary scientific contribution of this work. Accordingly, we focus on the proposed methodology rather than extensive hyperparameter optimization.

\subsubsection{Datasets}
We train RoCoMark, RoSeMary, and SrcMarker using the Python subset of CodeSearchNet \cite{husain2019codesearchnet}, with 25,000 Python functions for training and 2,500 for validation. To assess their generalization to unseen code, we evaluate the models on four datasets: HumanEval \cite{chen2021evaluating} (164 samples), MBPP \cite{austin2021program} (974 samples), EvalPlus \cite{liu2023your} (542 samples), and APPS \cite{hendrycks2021measuring}, for which we use a test subset of 5,000 samples.

\subsubsection{Baselines}
We select SWEET \cite{lee2024wrote}, SrcMarker \cite{yang2024srcmarker}, and RoSeMary \cite{zhang2025robust} as baseline methods. To ensure a fair comparison, RoCoMark, RoSeMary, and SrcMarker are trained on the same CodeSearchNet Python split and evaluated using identical watermark lengths, test data, attack samples, detection thresholds, and metric computation protocols. For each method, a single fixed checkpoint is used across all evaluation settings, including clean code, variable renaming, and LLM-based rewriting, thereby avoiding dataset-specific or attack-specific model selection. Each method retains its original watermark embedding mechanism to avoid inadvertently altering its intended capabilities. SWEET, in contrast, is a zero-bit generative watermarking method that does not support multi-bit message embedding. Therefore, $\operatorname{BitACC}$ and $\operatorname{MsgACC}$ are not applicable to SWEET, and we compare it with the other methods only in terms of $\operatorname{PassRate}$, $\operatorname{AUROC}$, $\operatorname{TPR}$, and $\operatorname{FPR}$. These evaluation metrics are defined in the following.

\subsubsection{Evaluation Metrics}
We evaluate the model performance along three dimensions: functionality preservation, message recovery, and watermark detection. We measure functionality preservation by using $\operatorname{PassRate}$. Let $\operatorname{Pass}(\mathcal{C})\in\{0,1\}$ indicate whether code $\mathcal{C}$ passes all test cases of corresponding sample. For a test set containing $N$ samples, $\operatorname{PassRate}$ is defined as
\begin{equation}
	\operatorname{PassRate}=
	\frac{1}{N}
	\sum_{i=1}^{N}
	\operatorname{Pass}\left(\mathcal{C}^{(i)}_w\right)
	\times 100\%.
\end{equation}
Here, $\mathcal{C}^{(i)}$ and $\mathcal{C}^{(i)}_w$ denote the original and watermarked code for $i$-th sample. For RoCoMark, RoSeMary, and SrcMarker, $\operatorname{PassRate}$ gives the proportion of watermarked code samples that pass all associated tests, considering only samples whose original code passes the tests. For SWEET, $\mathcal{C}^{(i)}$ and $\mathcal{C}^{(i)}_w$ denote the unwatermarked and watermarked code, respectively, and $\operatorname{PassRate}$ is computed over corresponding evaluation samples whose unwatermarked code passes all tests.

Message recovery is evaluated by $\operatorname{BitACC}$ and $\operatorname{MsgACC}$. Let $N_{+}$ be the number of marked code samples included in the evaluation, and let $\mathcal{C}^{(i)}_s$ and $\mathbf{m}^{(i)}$ denote the $i$-th evaluated code sample and its corresponding embedded message, respectively. For evaluations on watermarked code, $\mathcal{C}^{(i)}_s=\mathcal{C}^{(i)}_w$, whereas for rewriting-attack evaluations, $\mathcal{C}^{(i)}_s=\mathcal{C}^{(i)}_{rw}$. Using the matching-bit count $\mathcal{K}(\mathcal{C}^{(i)}_s;\mathbf{m}^{(i)})$ defined in Eq. (\ref{eq:k}), we have
\begin{equation}
	\left\{
	\begin{aligned}
		&\mathrm{BitACC} = \frac{1}{N_{+}L} \sum_{i=1}^{N_{+}} \mathcal{K}\!\left(\mathcal{C}^{(i)}_s; \mathbf{m}^{(i)}\right),\\
		&\mathrm{MsgACC} = \frac{1}{N_{+}} \sum_{i=1}^{N_{+}} I\!\left[ \mathcal{K}\!\left(\mathcal{C}^{(i)}_s; \mathbf{m}^{(i)}\right) = L \right].
	\end{aligned}
	\right.
\end{equation}
$\operatorname{BitACC}$ shows the average bit-wise recovery accuracy across all evaluated samples, while $\operatorname{MsgACC}$ gives the proportion of samples for which the entire message is recovered correctly. Both metrics are omitted for SWEET, as it does not support multi-bit message embedding.

We evaluate watermark detection according to $\operatorname{TPR}$, $\operatorname{FPR}$, and $\operatorname{AUROC}$. Positive samples are watermarked code samples, where $\mathcal{C}^{(i)}_s=\mathcal{C}^{(i)}_w$ under normal evaluation and $\mathcal{C}^{(i)}_s=\mathcal{C}^{(i)}_{rw}$ under LLM-rewriting attacks. Negative samples are the corresponding original, unwatermarked code samples $\mathcal{C}^{(i)}$. Let $N_+$ and $N_-$ denote the numbers of positive and negative samples, respectively. TPR and FPR are therefore defined as
\begin{equation}
\left\{
\begin{aligned}
    &\mathrm{TPR} = \frac{1}{N_+} \sum_{i=1}^{N_+} I\!\left[ \mathrm{WMarkDetection}(\mathcal{C}^{(i)}_s, \mathbf{m}^{(i)}) \geq \epsilon \right],\\
    &\mathrm{FPR} = \frac{1}{N_-} \sum_{i=1}^{N_-} I\!\left[ \mathrm{WMarkDetection}(\mathcal{C}^{(i)}, \mathbf{m}^{(i)}) \geq \epsilon \right].
\end{aligned}
\right.
\end{equation}
Here, $\mathbf{m}^{(i)}$ represents the target message associated with the corresponding code sample, and $\epsilon$ is the detection threshold. Although negative samples contain no watermark, their detection scores are computed against the same preset target messages to quantify false positives arising by chance. $\operatorname{TPR}$ measures the proportion of positive samples correctly identified as watermarked, whereas $\operatorname{FPR}$ measures the proportion of negative samples incorrectly classified as watermarked.

$\operatorname{AUROC}$ is defined as the area under the receiver operating characteristic (ROC) curve \cite{fawcett2006introduction}, i.e.,
\begin{equation}
\mathrm{AUROC}
=
\int_0^1 \mathrm{TPR}(u)~\text{d}u,
\end{equation}
where $u$ is the false positive rate as the detection threshold varies. $\operatorname{AUROC}$ quantifies the overall discriminative ability of a detector across different thresholds. For SWEET, we use its original zero-bit detection score to construct the ROC curve; the interpretation of $\operatorname{AUROC}$ remains unchanged.

Under the 4-bit setting used in this paper, the hard detection criterion is satisfied only when all four bits match the target message. Consequently, when $\operatorname{TPR}$ and $\operatorname{MsgACC}$ are computed over the same set of positive samples using the same message-matching criterion, their values coincide.

\subsubsection{Attacks}
We consider two representative attacks: variable renaming and LLM-powered code rewriting attacks. Both aim to disrupt watermark detection while preserving the tested functionality and public interface of the code.

\textbf{Variable-renaming.} The attacker randomly selects modifiable variables in the watermarked code and renames them without knowing which variables carry watermark information \cite{yang2024srcmarker,zhang2025robust}. VA0, VA10, VA30, and VA50 represent target renaming ratios of 0\%, 10\%, 30\%, and 50\%, respectively. For any nonzero target ratio, at least one modifiable variable is selected whenever available. Candidate identifiers are collected from the CodeSearchNet Python corpus and filtered to ensure validity and uniqueness \cite{husain2019codesearchnet}. The renaming procedure prioritizes candidate identifiers that are semantically similar to the original names and consistent with their naming styles. To preserve code validity, Python keywords, built-in names, existing identifiers, and conflicting candidates are excluded. Each selected variable is renamed consistently across its declaration and all corresponding references.

\textbf{LLM-powered code rewriting.} We employ GPT-5.5, DeepSeek-V4-Flash, MiMo-V2.5-Pro, GLM-5, and Claude-Sonnet-4.5 to rewrite watermarked code. For each model, we use four prompts with different levels of rewriting strength. These prompts instruct the models to modify variables, expressions, control flow, loop structures, or return paths while preserving the public interface and tested functionality. A rewritten code sample is included in the evaluation only if it passes syntax, interface, and functional correctness checks. TPR, BitACC, and MsgACC are evaluated on valid rewritten samples, while the corresponding unwatermarked code samples serve as negative samples for FPR and AUROC.

All experiments were conducted on a server equipped with eight NVIDIA GeForce RTX 3090 GPUs, each with 24 GB of memory. The experimental environment was configured with CUDA 12.4, PyTorch 2.6.0, and Python 3.11.14. To facilitate reproducibility, we will publicly release the simulation code\footnote[7]{URL: \url{https://github.com/qqlzyhello/RoCoMark}} to provide access to the implementation details.

\subsection{Watermark Detectability and Functionality Preservation}

\begin{figure*}[!htbp]
\centering
\includegraphics[width=0.8\textwidth]{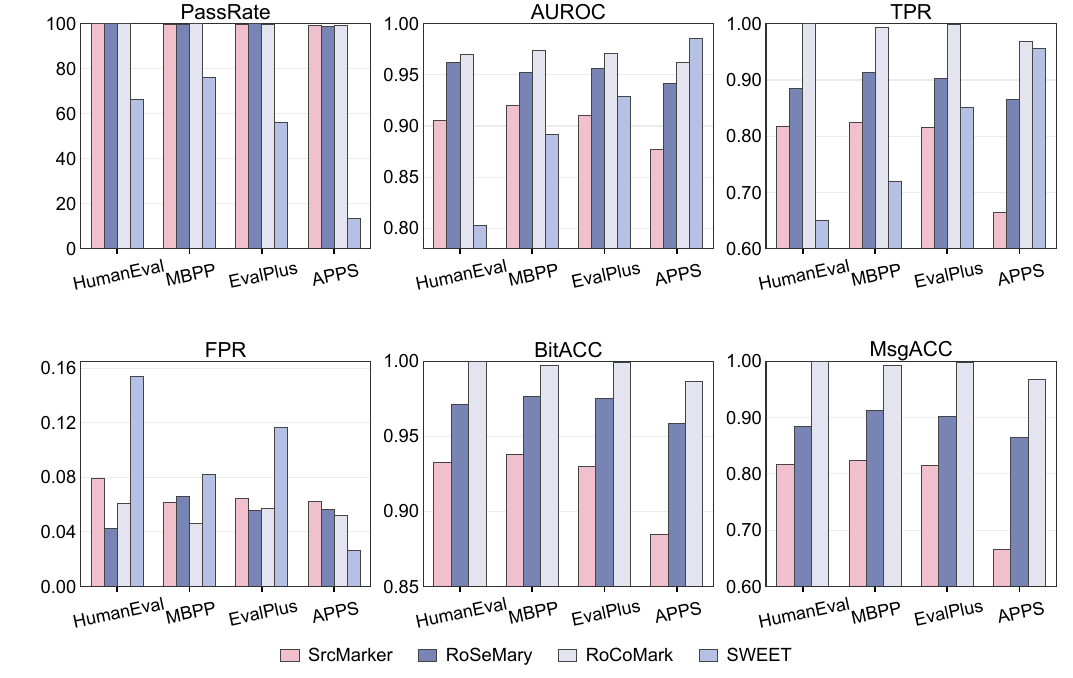}
\caption{Watermark detection and functionality-preservation performance of RoCoMark under the normal watermarking setting on HumanEval \cite{chen2021evaluating}, MBPP \cite{austin2021program}, EvalPlus \cite{liu2023your}, and APPS \cite{hendrycks2021measuring}, compared with the inference-time zero-bit watermarking method SWEET \cite{lee2024wrote} and the multi-bit code watermarking methods SrcMarker \cite{yang2024srcmarker} and RoSeMary \cite{zhang2025robust}.}
\label{fig:figure4}
\end{figure*}

Fig.~\ref{fig:figure4} compares the four methods in terms of functionality preservation, message recovery, and watermark detection under no-attack conditions. SrcMarker, RoSeMary, and RoCoMark all achieve $\operatorname{PassRate}$ close to 100\% across the four test sets, confirming that post-processing watermarking based on functionality-preserving transformations rarely compromises code correctness. In contrast, SWEET achieves lower $\operatorname{PassRate}$, with a more pronounced decline on challenging datasets such as APPS. This gap stems from the fundamental difference in embedding strategy. SWEET biases token probabilities during generation, and watermark-guided sampling can disrupt the subsequent generation trajectory. By contrast, the three post-processing methods apply constrained transformations only to already-generated, functionally correct code, making them inherently less intrusive.

For watermark detection, RoCoMark achieves high AUROC and TPR across all test sets while maintaining low FPR. This demonstrates that its detection capability generalizes from the CodeSearchNet Python training corpus to unseen code from diverse sources and difficulty levels. SWEET achieves competitive detection performance on some datasets but at the cost of reduced functionality preservation, reflecting an inherent trade-off between detectability and code correctness in generative watermarking.
\subsection{Robustness under Rewriting Attacks}
\begin{figure}[!htbp]
\centering
\includegraphics[width=0.8\columnwidth]{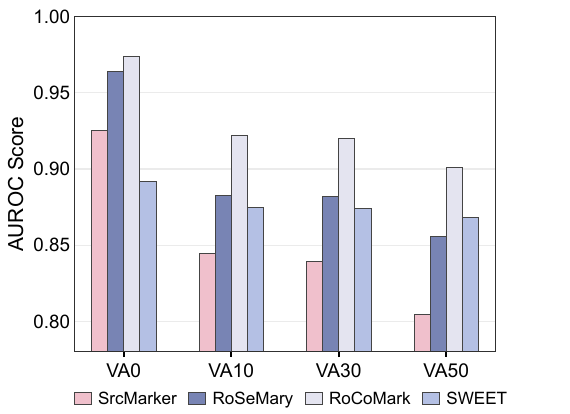}
\caption{Watermark-detection performance under the variable-renaming attack. VA0 denotes watermarked code with no variables replaced, whereas VA10, VA30, and VA50 denote settings in which 10\%, 30\%, and 50\% of the modifiable variables are replaced, respectively. All methods are evaluated on the same 974 MBPP samples using a shared detection threshold.}
\label{fig:figure5}
\end{figure}

\begin{figure}[!htbp]
\centering
\includegraphics[width=\columnwidth]{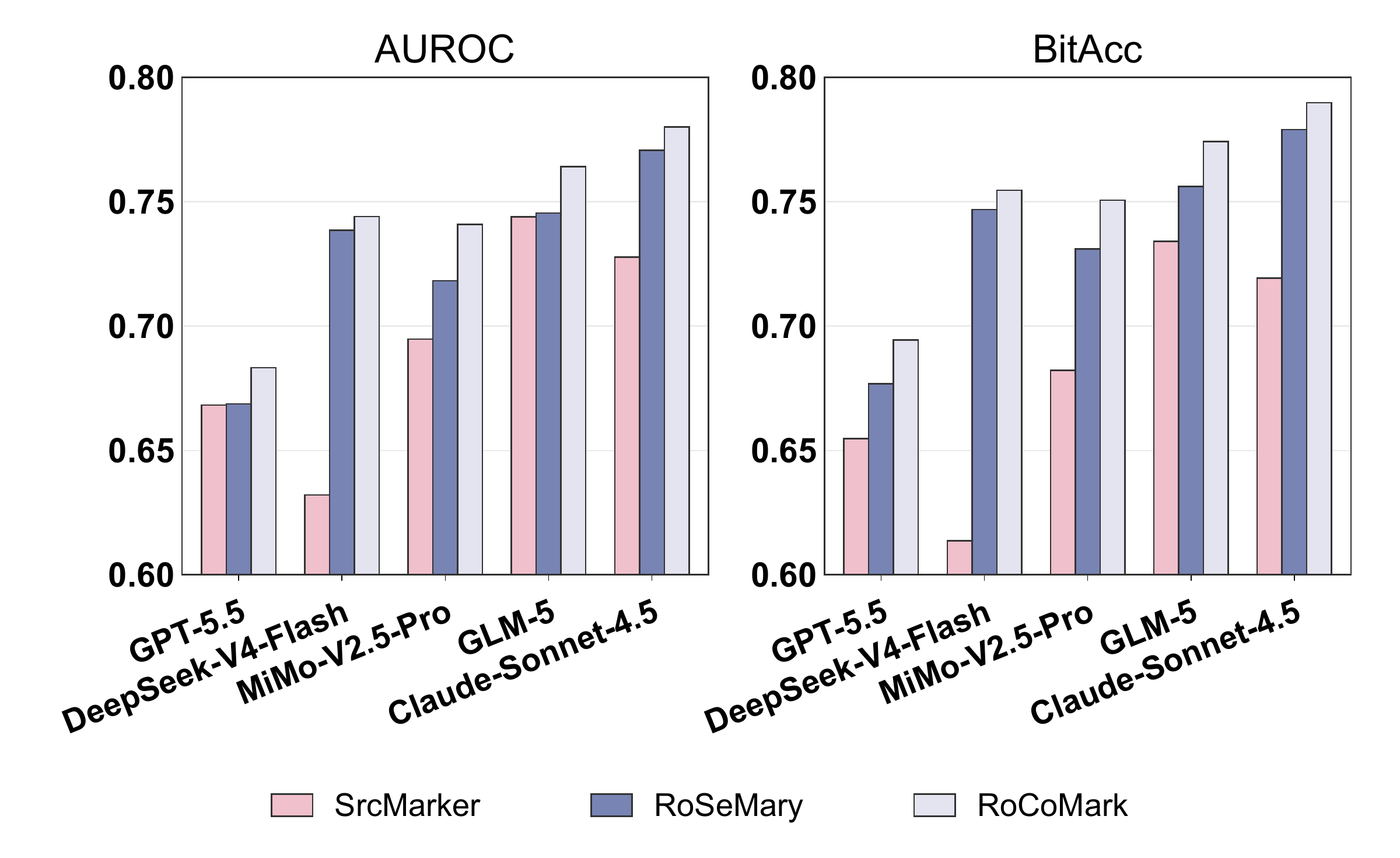}
\caption{Watermark detection performance under code rewriting attacks by different LLMs. The experiment employs GPT-5.5, DeepSeek-V4-Flash, MiMo-V2.5-Pro, GLM-5, and Claude-Sonnet-4.5 to generate functionality-preserving rewrites. Only valid rewritten code that passes syntax checks, function-signature checks, and functional tests is included in the evaluation.}
\label{fig:figure6}
\end{figure}
Fig.~\ref{fig:figure5} compares the AUROC of the four methods as the variable-renaming ratio increases. All methods show a decline in AUROC as the ratio rises from VA0 to VA50, confirming that variable names carry watermark-discriminative information. RoCoMark consistently achieves the highest AUROC across all renaming levels and exhibits a smaller decline than SrcMarker and RoSeMary. Even at VA50, its AUROC remains around 0.90, whereas those of SrcMarker and RoSeMary decline substantially more. This resilience suggests that RoCoMark does not rely solely on a limited set of variable names. Instead, its syntactic watermark carriers and robustness training exploit structural and contextual cues that persist after renaming. SWEET also shows a decline in performance, indicating that changes to variable tokens perturb its watermark statistics, even though zero-bit generative watermarking lacks explicit message encoding.

Fig.~\ref{fig:figure6} compares the detection performance of SrcMarker, RoSeMary, and RoCoMark under code-rewriting attacks by five LLMs. For a fair comparison, all attacks use the same watermarked code, rewriting prompts, and validity checks. Detection metrics are computed only for rewritten code that passes syntax, interface, and functional checks. The AUROC and BitAcc of the three multi-bit watermarking methods varies across attacker LLMs. GPT-5.5 and DeepSeek-V4-Flash produce rewrites that disrupt the watermark more aggressively, whereas GLM-5 and Claude-Sonnet-4.5 better preserve watermark information. This result indicates that the strength of an LLM-based rewriting attack depends not only on the prompt but also on the LLM employed.

Although attacker LLMs produce distinct rewrite distributions, RoCoMark consistently achieves the highest AUROC across all five attacks, with an average AUROC of 74.2\% and a BitACC of 75.3\%, indicating that its robustness gains generalize across rewriting models. Under the MiMo-V2.5-Pro attack, RoCoMark attains an AUROC of 0.7408 and a TPR of 0.4638, outperforming RoSeMary (0.7182, 0.4062) and SrcMarker (0.6948, 0.3617), while its FPR drops to 0.0463. These results confirm that the rewrite-stable watermark representations learned in the Threat Profiling module, together with the dynamic adversarial game between the watermarking model and the attacker in Strategy Improvement, enhance watermark detectability under global structural rewriting and suppress false positives on unwatermarked code.

\subsection{Ablation Studies and Analysis}
\begin{figure}[!htbp]
\centering
\includegraphics[width=\columnwidth]{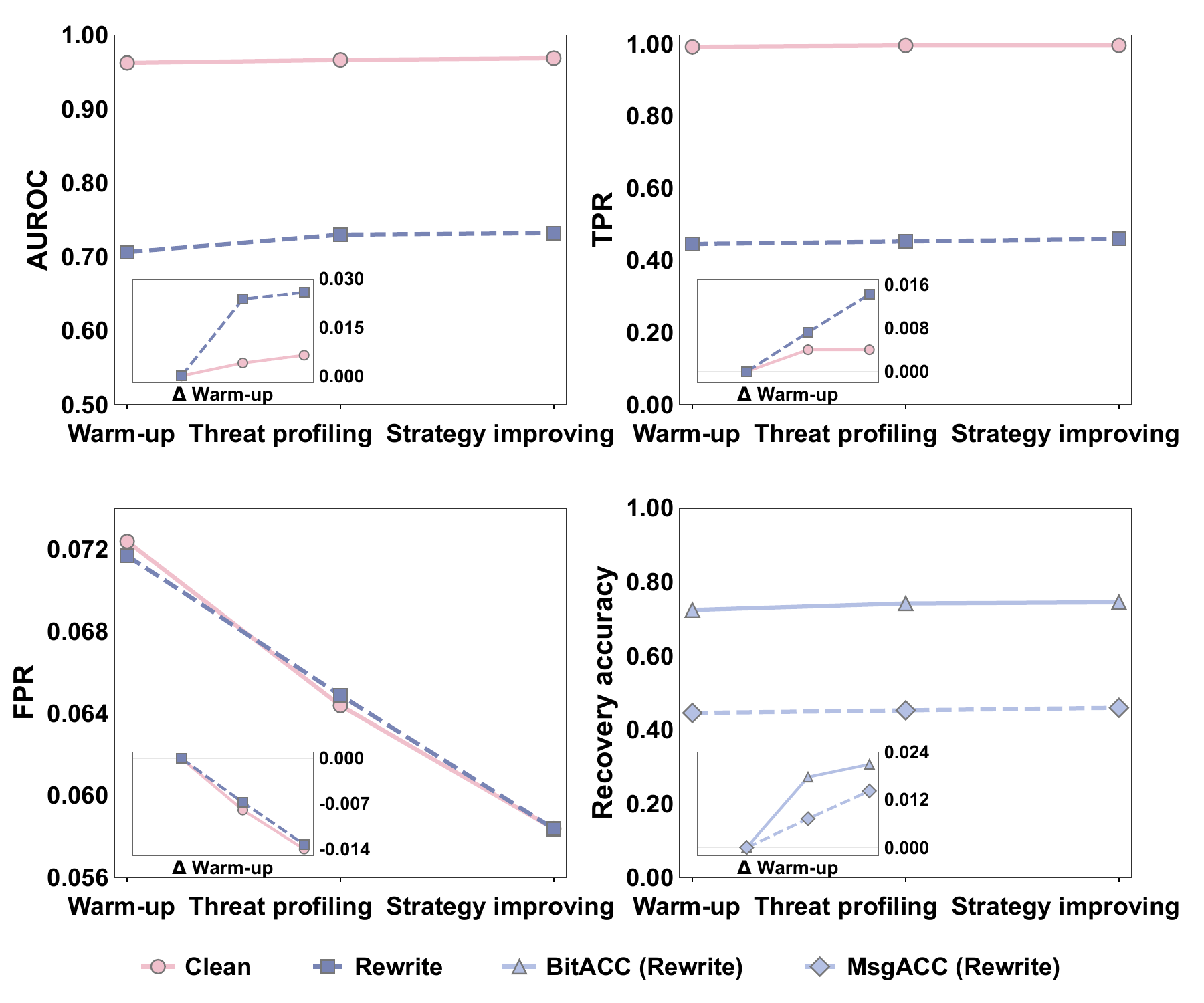}
\caption{Ablation study on the three training modules of RoCoMark. It denotes the improvement of each remaining module over the Warm-up module.}
\label{fig:figure7}
\end{figure}

To assess the contributions of the three training modules, we evaluate Warm-up, Threat Profiling, and Strategy Improvement sequentially, as shown in Fig.~\ref{fig:figure7}. All models are evaluated on the same MBPP test set and using the same set of rewrites generated by MiMo-V2.5-Pro.

\begin{table}[!t]
\caption{Watermark-detection performance during iterative optimization.}
\label{tab:table1}
\centering
\setlength{\tabcolsep}{4pt}
\renewcommand{\arraystretch}{1.15}
\begin{tabular*}{0.95\columnwidth}{@{\hspace{4pt}\extracolsep{\fill}}ccccc@{\hspace{4pt}}}
\toprule
 & Clean & \multicolumn{3}{c}{Rewrite} \\
\cmidrule(lr){2-2}\cmidrule(lr){3-5}
Round & AUROC & AUROC & TPR & FPR \\
\midrule
0 & 0.9688 & 0.7320 & 0.4594 & 0.0584 \\
1 & 0.9714 & 0.7341 & \textbf{0.4664} & 0.0519 \\
2 & \textbf{0.9742} & \textbf{0.7408} & 0.4638 & \textbf{0.0463} \\
3 & 0.9732 & 0.7403 & 0.4638 & 0.0485 \\
\bottomrule
\end{tabular*}
\end{table}

The Threat Profiling module increases the Rewrite AUROC from 0.7061 to 0.7299 and reduces the Rewrite FPR from 0.0716 to 0.0648. These results demonstrate that contrastive learning and anchor-based distillation enhance the stability of encoder representations across rewrites. Adding Strategy Improvement further increases the Rewrite AUROC and Rewrite TPR to 0.7320 and 0.4594, respectively, while reducing the Rewrite FPR to 0.0584. Notably, all three stages maintain a $\operatorname{PassRate}$ of 99.80\%, confirming that robustness improvements do not come at the cost of code functionality.

RoCoMark supports multiple rounds of Strategy Improvement optimization initialized from the Threat Profiling model. Table~\ref{tab:table1} shows how watermark-detection performance on rewritten code evolves during iterative optimization. The model improves over the first two rounds. The Rewrite AUROC increases from 0.7320 at step 0 to 0.7408 at round 2, while the FPR decreases from 0.0584 to 0.0463. Further training yields no additional gains and instead leads to a slight decline, suggesting that excessive optimization on a fixed rewriting distribution may cause the model to overfit to the attack samples used in training.

\begin{table}[!t]
\caption{Effectiveness of four progressively strengthened rewriting prompts against RoCoMark (Warm-up only)}
\label{tab:table2}
\centering
\setlength{\tabcolsep}{4pt}
\renewcommand{\arraystretch}{1.15}
\begin{tabular*}{0.75\columnwidth}{@{\hspace{4pt}\extracolsep{\fill}}ccc@{\hspace{4pt}}}
\toprule
Round & BitACC & MsgACC \\
\midrule
0 & 0.7935 & 0.5669 \\
1 & 0.7611 & 0.5137 \\
2 & 0.7337 & 0.4417 \\
3 & 0.6356 & 0.3090 \\
\bottomrule
\end{tabular*}
\end{table}

Table~\ref{tab:table2} compares the effectiveness of four rewriting prompts against RoCoMark (Warm-up only). The resulting attack strengths vary substantially across prompts. The four rewriting prompts affect both watermark-detection and message-recovery performance. As the scope of permitted modifications broadens, AUROC, TPR, BitACC, and MsgACC generally decrease. Prompt 1 primarily makes localized changes to variables or expressions, resulting in a relatively weak attack. Prompt 4 encourages simultaneous changes to variable names, control flow, loop structures, intermediate variables, and return paths. This prompt reduces AUROC to 0.6143, making it the most watermark-disruptive of the four.

\begin{table}[!t]
\caption{The trade-off analysis between watermark capacity and message-recovery accuracy.}
\label{tab:table3}
\centering
\setlength{\tabcolsep}{4pt}
\renewcommand{\arraystretch}{1.15}
\begin{tabular*}{0.85\columnwidth}{@{\hspace{4pt}\extracolsep{\fill}}cccc@{\hspace{4pt}}}
\toprule
Bits & SrcMarker & RoSeMary & RoCoMark \\
\midrule
2 & 0.9784 & \textbf{0.9923} & 0.9877 \\
4 & 0.9203 & 0.9520 & \textbf{0.9742} \\
6 & 0.8403 & 0.8535 & \textbf{0.9868} \\
8 & 0.7844 & 0.8201 & \textbf{0.9122} \\
\bottomrule
\end{tabular*}
\end{table}

We further vary the watermark length on MBPP to evaluate the effect of message capacity on extraction accuracy. To ensure architectural comparability, we train a separate base watermarking model for each bit length. We refer to this evaluation as the RoCoMark capacity experiment. BitACC decreases for all three methods as the watermark length increases (see Table~\ref{tab:table3}), because the available transformation space must encode more information. RoCoMark nonetheless achieves BitACC values of 0.9742, 0.9868, and 0.9122 at 4, 6, and 8 bits, respectively, compared with 0.8201 for RoSeMary and 0.7844 for SrcMarker at 8 bits. These results show that RoCoMark’s bitwise recovery remains comparatively stable as capacity increases, although further increases remain constrained by the available transformation space.

\subsection{Case Studies}
\begin{figure}[!t]
  \centering
  \includegraphics[width=0.8\columnwidth]{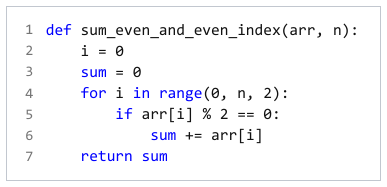}
  \caption{Original implementation of the running example, which sums the  even-valued elements at even indices.}
  \label{fig:code1}
\end{figure}

Fig.~\ref{fig:code1} shows the original implementation of the running example, which returns the sum of the even-valued elements at even indices. Each method embeds a 4-bit watermark. DeepSeek-V4-Flash then rewrites the watermarked code using Prompt~3, which permits changes to parameter and local-variable names and encourages rewriting loops, conditionals, intermediate variables, and accumulation.

\begin{figure*}[!t]
  \centering
  \includegraphics[width=0.9\textwidth]{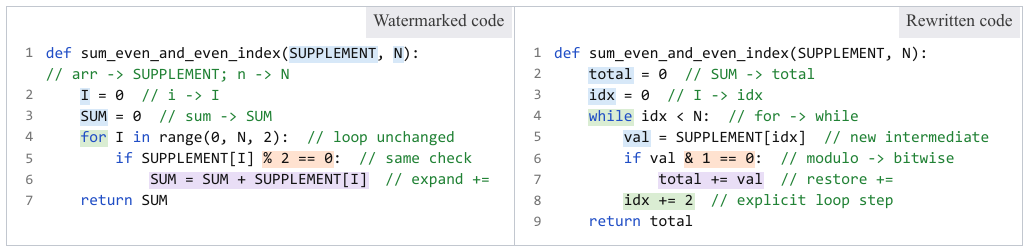}
  \caption{SrcMarker-watermarked code (left) and the corresponding LLM-rewritten version (right).}
  \label{fig:code2}
\end{figure*}

As shown in Fig.~\ref{fig:code2}, SrcMarker changes the parameter name, identifier style, and accumulation form. After rewriting by DeepSeek-V4-Flash, the variables are renamed, the loop is converted to a \texttt{while} loop, an intermediate variable is introduced, the parity check becomes bitwise, and augmented assignment is used. Thus, all surface-syntax carriers introduced by SrcMarker are altered.

\begin{figure*}[!t]
  \centering
  \includegraphics[width=0.9\textwidth]{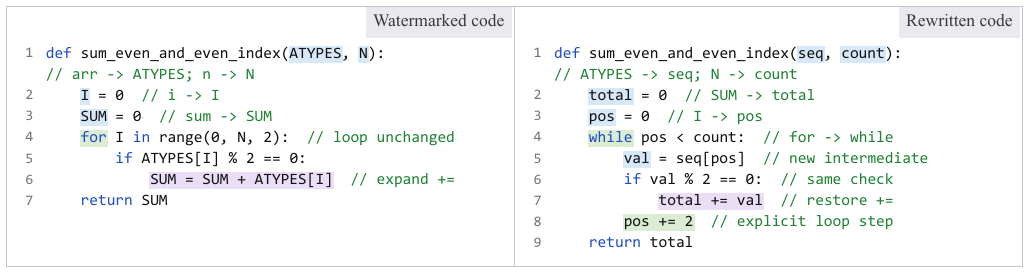}
  \caption{RoSeMary-watermarked code (left) and the corresponding LLM-rewritten version (right).}
  \label{fig:code3}
\end{figure*}

As shown in Fig.~\ref{fig:code3}, RoSeMary uses uppercase identifiers. The rewrite renames the parameters and local variables, converts the loop to a \texttt{while} loop, introduces an intermediate variable, and changes the accumulation form. Its identifier pattern and local structure are therefore not fully preserved.

\begin{figure*}[!t]
  \centering
  \includegraphics[width=0.9\textwidth]{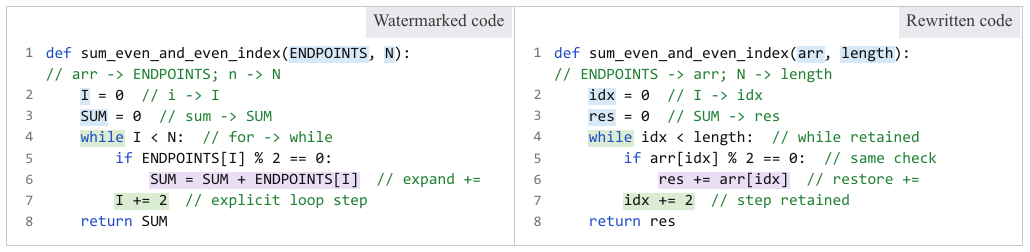}
  \caption{RoCoMark-watermarked code (left) and the corresponding LLM-rewritten version (right).}
  \label{fig:code4}
\end{figure*}

As shown in Fig.~\ref{fig:code4}, RoCoMark uses uppercase identifiers and a \texttt{while} loop. The rewrite renames local variables and formal parameters and uses augmented assignment, but retains the \texttt{while}-loop structure and modulo-based condition without introducing additional intermediate variables.

These transformations destroy the variable-naming and syntactic-form carriers used by SrcMarker and RoSeMary. RoCoMark instead couples its carrier to the program’s intrinsic computational logic: despite changes to the function name, parameters, local variables, and accumulation expression, the core loop structure remains intact and the complete 4-bit message is recovered, as reported in Table~\ref{tab:successful-watermark-detection-case}.

\begin{table}[!htbp]
    \centering
    \caption{A case of watermark detection under an LLM-powered code rewriting attack}
    \label{tab:successful-watermark-detection-case}
    \footnotesize
    \setlength{\tabcolsep}{3pt}
    \renewcommand{\arraystretch}{1.15}
    \begin{tabular}{lcccc}
        \hline
        \textbf{Method}
        & \makecell{$\mathbf{m}$}
        & \makecell{$\hat{\mathbf{m}}_{\mathcal{C}_w}$}
        & \makecell{$\hat{\mathbf{m}}_{\mathcal{C}_{rw}}$}
        & \makecell{\textbf{Number of Matching Bits}} \\
        \hline
        SrcMarker
        & \texttt{0011}
        & \texttt{0011}
        & \texttt{1011}
        & \texttt{3/4} \\
        
        RoSeMary
        & \texttt{0011}
        & \texttt{0011}
        & \texttt{0101}
        & \texttt{2/4} \\
        
        RoCoMark
        & \texttt{0011}
        & \texttt{0011}
        & \texttt{0011}
        & \texttt{4/4} \\
        \hline
    \end{tabular}
\end{table}

\section{Conclusion and Discussion}

RoCoMark formulates a multi-round minimax game between a defender and a set of attackers. It comprises three training modules, including a Reflexion-based adversarial attacker designed to improve the robustness of source-code watermarks against LLM-driven, functionality-preserving code rewriting. In the Threat Profiling module, contrastive learning and knowledge distillation stabilize encoder representations across rewrites. In the Strategy Improvement module, the attacker and the watermarking model co-evolve through an adversarial game that continually exposes the model to stronger attacks, thereby progressively improving its robustness to rewriting. Experiments across four benchmarks and five LLMs validate the effectiveness of RoCoMark and its ability to preserve functionality. Future work will explore more efficient adversarial sample generation, novel watermark encodings, and robustness to cross-language rewriting.


\bibliographystyle{IEEEtran}
\bibliography{references}

\includepdf[pages=-]{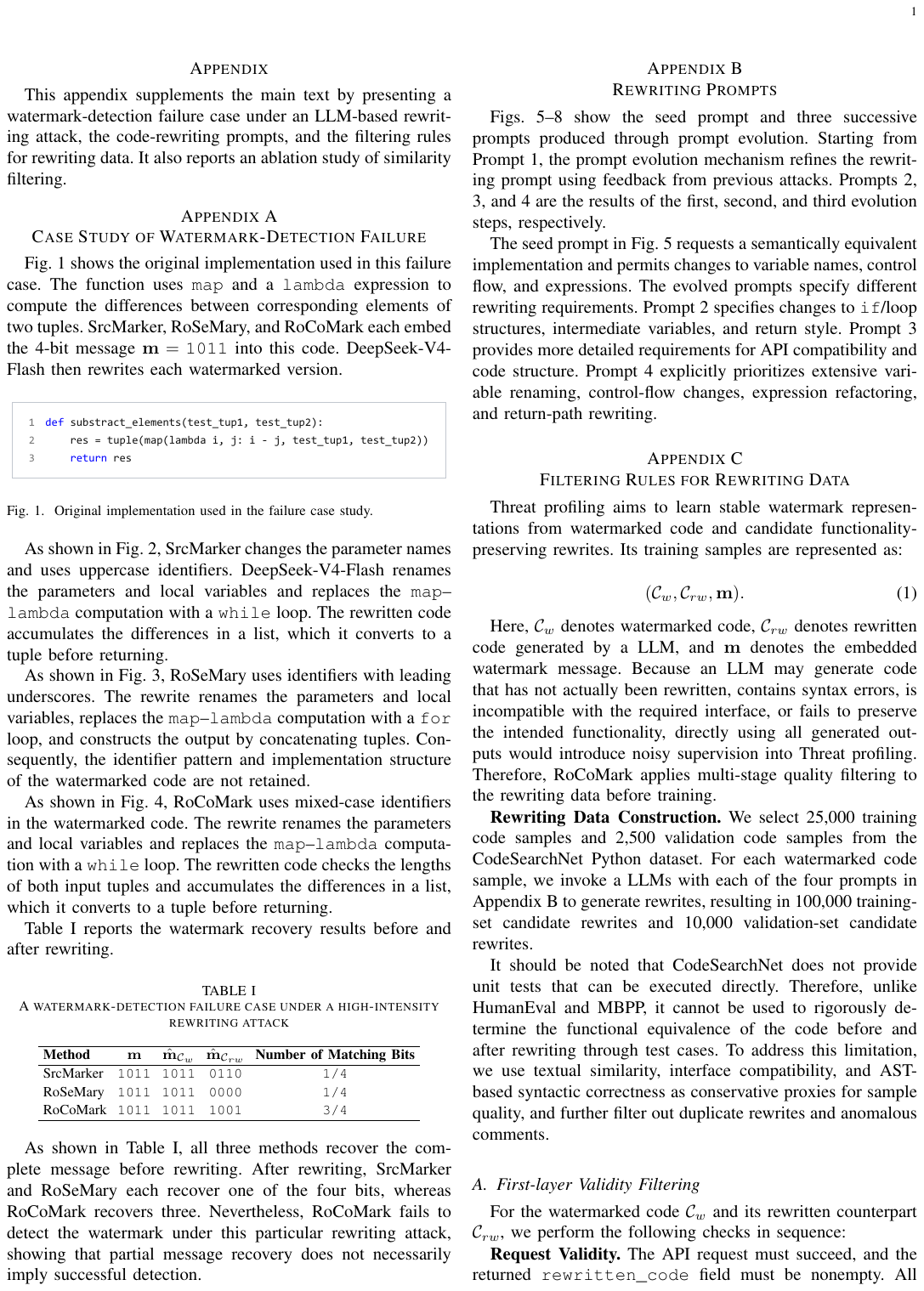} 

\end{document}